\documentclass[conference]{IEEEtran}
\usepackage{makeidx}
\usepackage{color}
\usepackage{cite}
\usepackage[hyphens]{url}
\usepackage{wrapfig}
\makeatletter
\def\url@leostyle{%
  \@ifundefined{selectfont}{\def\UrlFont{\sf}}{\def\UrlFont{\small\ttfamily}}}
\makeatother
\usepackage{graphics}
\usepackage{subfigure}
\usepackage[ruled]{algorithm2e}
\usepackage{array, varwidth}
\usepackage{amssymb,amsmath,epsfig}

\newcommand{\nop}[1]{}
\newtheorem{thm}{Theorem}
\newtheorem{lem}[thm]{Lemma}
\newtheorem{remark}{Remark}
\newtheorem{cor}{Corollary}

\begin{document}
%
\title{2-Edge-Connectivity and 2-Vertex-Connectivity of an Asynchronous
Distributed Network}

\author{Abusayeed Saifullah\\
Department of Computer Science, Wayne State University
}

%
%
%

\maketitle
\pagestyle{plain}
\setcounter{page}{1} 
 \thispagestyle{plain}
\begin{abstract}

Self-stabilization for non-masking fault-tolerant distributed system has received
considerable research interest over the last decade. In this
paper, we propose a self-stabilizing algorithm for
2-edge-connectivity and 2-vertex-connectivity of an asynchronous
distributed computer network. It is based on a self-stabilizing
depth-first search, and is not a composite algorithm in the sense 
that it is not composed of a number of self-stabilizing algorithms that run concurrently.
The time and space complexities of the algorithm are the same as
those of the underlying self-stabilizing depth-first search
algorithm which are $O(dn\Delta )$ rounds and $O(n\log \Delta )$
bits per processor, respectively, where $\Delta (\le n)$ is an
upper bound on the degree of a node, $d (\le n)$ is the diameter
of the graph, and $n$ is the number of nodes in the network.

\end{abstract}

\IEEEpeerreviewmaketitle

\vspace{2ex} \noindent {\bf\normalsize KEY WORDS}\newline
{Distributed system,   fault-tolerance, self-stabilization,
depth-first search tree, bridge, articulation point,
bridge-connected component.}

\section{Introduction}\label{introduction}A \emph{distributed system} is a set of processing elements or state
machines interconnected by a network of some fixed topology.
Distributed systems are exposed to constant changes of their
environment and the design of such systems is quite complex, in
part due to unpredictable faults. Implicit in the notion of
\emph{fault} is the specification of what constitutes the correct
state of the system. A \emph{transient fault} is an event that may
change the state of a system by corrupting the local states of the
machines. The property of \emph{self-stabilization} can recover
the system from transient faults and represents a departure from
previous approaches to fault tolerance.

The notion of \emph{self-stabilization} was first proposed by
Dijkstra~\cite{dijkstra, belated}. A system is
\emph{self-stabilizing} if, starting at any state, possibly
illegitimate, it eventually converges to a legitimate state in
finite time \cite{ijfcs, ispa}. A self-stabilizing system is capable of tolerating
any unexpected transient fault without being assisted by any
external agent. Regardless of the initial state, it can reach a
legitimate global state in finite time and can remain so
thereafter unless it experiences any subsequent fault. In this
paper, we propose a simple self-stabilizing algorithm for
detecting the bridges, articulation points, and bridge-connected
components of an asynchronous distributed network. When a
distributed system is modelled as an undirected connected graph,
an edge is called a \emph{bridge} if its removal disconnects the
graph whereas an \emph{articulation point} is a node whose removal
disconnects the graph. A maximal component without any bridge of
the graph is called a \emph{bridge-connected component}.
Bridge-connectivity (2-edge-connectivity) and biconnectivity
(2-vertex-connectivity) call for considerable attention in graph
theory since these properties represent the extent to which a
graph is connected \cite{cats}. In distributed systems, these properties
represent the reliability of the network in presence of link or
node failures. Moreover, when communication links are expensive,
these properties play a vital role to minimize the communication
cost.

Several self-stabilizing algorithms for 2-edge-connectivity and
2-vertex-connectivity are available. The algorithm
in~\cite{chaudhurybcc} can find the bridge-connected components by
assuming the existence of a depth-first search spanning tree of
the system. This algorithm stabilizes in two phases and, for a
system with $n$ processors, each phase requires \emph{O}$(n^2)$
moves to reach a legitimate configuration by assuming that the
preceding phase has stabilized. If a breadth-first search tree of
the network is known, then the algorithm in~\cite{karaatabridge}
can detect the bridges in \emph{O}$(n^3)$ moves and that
in~\cite{karaataarticulation} can detect the articulation points
in \emph{O}$(n^3)$ moves. The algorithm in~\cite{karaatabc} finds
the biconnected components in \emph{O}$(n^2)$ moves if a
breadth-first search tree and all the bridges of the network are
known. Each of the algorithms~\cite{chaudhurybcc,
karaataarticulation, karaatabc, karaatabridge} mentioned above
requires \emph{O}$(n\Delta\lg \Delta)$ bits per processor, where
$\Delta $ is an upper bound on the \emph{degree} of a processor.
The algorithm proposed by Devismes~\cite{devismes} uses a weaker
model (one that does not require every node to have a distinct
identifier) and can detect the cut-nodes and bridges in
\emph{O}$(n^2)$ moves if a depth-first search tree of the network
is known. This algorithm is memory efficient (\emph{O}$(n\lg
\Delta +\lg n)$ bits per processor) but does not find the
bridge-connected or biconnected components.

It is pointed out in~\cite{tsinss} that each of the aforementioned
algorithms is just one component of a composite algorithm  and
hence the time complexity presented is different from that of the
composite algorithm.  Since the algorithm must run concurrently
with a self-stabilizing spanning tree algorithm (which is another
component of the composite algorithm), when the last transient
fault had elapsed and the spanning tree algorithm has stabilized,
the processor may make redundant moves on the spanning tree
algorithm which could significantly lengthen the time that the
composite algorithm needs to stabilize. In the worst case, the
time complexity of the composite algorithm is the $product$ of the
time complexities of the algorithms that make up the composite
algorithm and is thus bounded below by that of the spanning tree
algorithm. Addressing all these issues,
 Tsin~\cite{tsinss} has shown how to incorporate
Tarjan's depth-first-search based algorithm for biconnectivity
into the self-stabilizing depth-first search algorithm of Collin
and Dolev~\cite{collin} to produce a self-stabilizing algorithm
for bridge-connectivity and biconnectivity. The time and space
complexities of the resulting algorithm are bounded above by those
of the depth-first search algorithm. Following this elegant
approach~\cite{tsinss}, our algorithm simplifies all existing
algorithms for bridge-connectivity and
biconnectivity~\cite{chaudhurybcc,devismes,karaataarticulation,karaatabc,karaatabridge}
by embedding the detection method of bridges and articulation
points in the self-stabilizing depth-first search algorithm of
Collin and Dolev~\cite{collin} and by avoiding any distributed
protocol composition. The proposed algorithm also determines all
the bridge-connected components since, upon stabilization of the
algorithm, all the nodes of the same component contain the same
identifier. The space complexity is also significantly improved in
our algorithm. The space requirement for
 each of the algorithms of~\cite{chaudhurybcc,karaataarticulation,karaatabc,karaatabridge} is \emph{O}$(n^2 \log (n))$ bits per processor for a
system with $n$ processors. This is due to the propagation of a
set of non-tree edges that bypass a tree edge in the depth-first
search spanning tree of the system. However, we show that passing
only the size of that set is sufficient for detecting all the
bridges and articulation points which substantially reduces the
size of the message. Specifically, the time complexity of our
algorithm is $O(dn\Delta)$ rounds and the space complexity for
every processor is \emph{O}($n \log\Delta$) bits. Note that the
space complexity for the self-stabilizing depth-first algorithm of
Collin and Dolev~\cite{collin} is \emph{O}($n \log\Delta$) bits
per processor and the time complexity is $O(dn\Delta)$ rounds. The
model we use is the same as that of Collin and
Dolev~\cite{collin}, which is weaker than that used
in~\cite{chaudhurybcc, chaudhuryfc, karaataarticulation,
karaatabc, karaatabridge}.

\section{Computational Model}\label{model}
The distributed system is represented by an undirected connected
 graph $G=(V,E)$. The set of nodes $V$ in $G$
represents the \emph{set of processors} $\{v_1,v_2, \cdots ,v_n\}$,
where $n$ is the total number of processors in the system, and $E$
represents the \emph{set of bidirectional communication links}
between two processors. We shall use the terms \emph{node} and
\emph{processor} (\emph{edge} and \emph{link}, respectively)
interchangeably throughout this paper. We assume that the graph is
bridgeless.

All the processors, except $v_1$, are anonymous. The processor
$v_1$ is a special processor and is designated as the \emph{root}.
For the processors $v_i$, $2\le i\le n$, the subscripts $2,\cdots,
n$ are used for ease of notation only and must not be interpreted
as identifiers. Two processors are \emph{neighboring} if they are
connected by a link. The processors run asynchronously and the
communication facilities are limited only between the neighboring
processors. Communication between the neighbors is carried out
using \emph{\textbf{shared communication registers}} (called
\emph{\textbf{registers}} throughout this paper). Each register is
\emph{serializable} with respect to \emph{read} and \emph{write}
operations.

Every processor $v_i$, $1\le i\le n$, contains a register.  A
processor can both read and write to its own register.  It can
also read the registers of the neighboring processors but cannot
write to those registers. The contents of the registers are
divided into \emph{fields}. Each processor $v_i$ orders its edges
by some arbitrary ordering $\alpha_i$. For any edge $e=(v_i,
v_j)$, $\alpha_i(j)$ ($\alpha_j(i)$, respectively) denotes the
\emph{edge index} of $e$ according to $\alpha_i$ ($\alpha_j$,
respectively). Furthermore, for every processor $v_i$ and any edge
$e=(v_i, v_j)$, $v_i$ knows the value of $\alpha_j(i)$.

We consider a processor and its register to be a single entity, thus
the \emph{state of a processor} fully describes the value stored in
its register, program counter, and the local variables. Let $\chi_i$
be the set of possible states of processor $v_i$.
 A \emph{\textbf{configuration}}  $c\in (\chi_1\times \chi_2\times \cdots \chi_n)$ of the
system is a \emph{vector} of states, one for each processor.
Execution of the algorithm proceeds in steps (or \emph{atomic}
steps) using \emph{\textbf{read/write atomicity}}. An
\emph{\textbf{atomic step}} of a processor consists of an internal
computation followed by either \emph{\textbf{read}} or
\emph{\textbf{write}}, but not both. Processor activity is managed
by a \emph{\textbf{scheduler}} (also called \emph{\textbf{daemon}}).
At any given configuration, the scheduler activates a single
processor which executes a single \emph{atomic} step.

An \emph{\textbf{execution}} of the system is an infinite sequence
of configurations $\Re=(c_0, c_1, \cdots, c_i, c_{i+1},$ $ \cdots
)$ such that for $i\ge 0$, $c_i\rightarrow c_{i+1}$ (called a
\emph{\textbf{single computation step}} ) denotes that
configuration $c_{i+1}$ can be reached from configuration $c_{i}$
by executing on step. A \emph{\textbf{fair execution}} is an
infinite execution in which every processor executes atomic steps
infinitely often. A \emph{\textbf{suffix}} of a sequence of
configurations $(c_0, c_1, \cdots, c_i, c_{i+1}, \cdots )$ is a
sequence  $(c_k, c_{k+1}, \cdots )$, where $k\ge 0$. The finite
sequence $(c_0, c_1, \cdots, c_{k-1} )$ is a
\emph{\textbf{prefix}} of the sequence of configurations. A
\emph{\textbf{task}} is defined by a set of executions, called
\emph{\textbf{legal executions}}. A distributed algorithm is
\emph{\textbf{self-stabilizing}} for a task if every fair
execution of the algorithm has a \emph{suffix} belonging to the
set of legal executions of that task. The time complexity of the
algorithm is expressed in terms of the number of
\emph{\textbf{rounds}}~\cite{dolev00}. The \emph{first round} of
an execution $\Re$ is the shortest prefix of $\Re$ in which every
processor executes at least one step. Let $\Re=\Re_1\Re_2$ such
that $\Re_1$ is the prefix consisting of the first $k$ rounds of
$\Re$. Then the ($k+1$)-th round of $\Re$ is the first round of
$\Re_2$.

\section{The Algorithm}\label{algorithm}
The algorithm uses the self-stabilizing depth-first search
algorithm of Collin and Dolev~\cite{collin} to construct a
depth-first search spanning tree. In the self-stabilizing
depth-first search algorithm of Collin and Dolev~\cite{collin},
every processor $v_i$ has a field, denoted by $path_i$, in its
register. At any point of time during the execution of the
algorithm, $path_i$ contains the sequence of indices of the links
on a path connecting the root $v_1$ with node $v_i$. The algorithm
uses a \emph{lexicographical order relation} $\prec$ on the path
representation and the \emph{concatenation} of any link with a
path is denoted by the operator $\oplus$. The root processor $v_1$
always writes $\bot$ in its $path_1$ field and, in the
lexicographical order relation,  $\bot$ is the \emph{minimal
character}. When a depth-first search tree is constructed in the
network, $path_i$ contains the smallest (with respect to the
lexicographical order $\prec$) path connecting $v_1$ with $v_i$.
The last links on the smallest paths of $v_i$, $i\ge 2$, form a
depth-first search tree, called the \emph{first depth-first search
tree}. Given that in the first depth-first search tree, a node
$v_j$ is an ancestor of a node $v_i$ if the smallest path of $v_i$
contains the smallest path of $v_j$, then, the node $v_j$ is an
ancestor of a node $v_i$ if $path_j$ is a prefix of $path_i$, i.e.
$(\exists s)(path_i=path_j\oplus s)$. If there exists a unique
neighbor $v_j$ of $v_i$ such that $path_i=path_j\oplus
 \alpha_j(i)$, then $v_j$ is the \emph{parent} of $v_i$.
The \emph{\textbf{degree of a processor}} $v_i$, denoted by
$\delta_i$, is the number of incident edges (links) on $v_i$. Once
 a depth-first search tree is constructed, at each processor $v_i$,
the type of each incident link ($v_i, v_j$) (or ($v_j, v_i$)) can
be determined by $path_i, path_j, \alpha_i(j)$, and $\alpha_j(i)$
in the following ways:

\begin{itemize}
    \item The link ($v_j,v_i$) is a \emph{\textbf{parent link}} if and only
    if $path_i=path_j\oplus \alpha_j(i)$;
    \item The link ($v_i,v_j$) is a \emph{\textbf{child link}} if and only
    if $path_j=path_i\oplus \alpha_i(j)$;
    \item The link ($v_i,v_j$) is an \emph{\textbf{outgoing non-tree edge}} (i.e. it is a non-tree link and $v_j$ is an ancestor of $v_i$) if and only
    if ($\exists s$)(($path_i=path_j\oplus
    s$)$\wedge$($s\not=\alpha_j$($i$))); The total number of
    outgoing non-tree edges incident on processor $v_i$ is denoted
    by $out_i$.
    \item The link ($v_j,v_i$) is an \emph{\textbf{incoming non-tree edge}} (i.e. it is a non-tree link and $v_j$ is a descendant of $v_i$) if and only
    if ($\exists s$)(($path_j=path_i\oplus
    s$)$\wedge$($s\not=\alpha_i$($j$))); The total number of incoming non-tree edges incident on processor $v_i$ is denoted
    by $in_i$.
    \end{itemize}

We omit the description of that part of the algorithm for
constructing a depth-first search tree $T$, as it is available
in~\cite{collin}. The idea underlying our algorithm is to count
the total number of non-tree edges that bypass a tree edge in $T$.
A non-tree edge ($v_k, v_l$) ($v_k$ is a descendant of $v_l$)
\textbf{\emph{bypasses}} a tree edge ($v_i, v_j$) ($v_i$ is the
parent of $v_j$) if and only if $v_k$ is a descendant of $v_j$
while $v_l$ is an ancestor of $v_i$. The total number of non-tree
edges bypassing the parent link of processor $v_i$ is denoted by
$count_i$. During the execution of the algorithm  this number is
propagated towards the root whereas in~\cite{chaudhurybcc,
karaatabridge, karaataarticulation, karaatabc}, for every node
$v_i$, the whole set of non-tree edges bypassing the parent link
of $v_i$ is calculated and routed towards the root. Since, in our
algorithm, only the cardinality of the set is propagated, the
message cost is drastically reduced. IN $T$, let $in_i$ and
$out_i$ be the number of incoming non-tree edges and the number of
outgoing non-tree edges, respectively, incident on $v_i$, and
$C_i$ be the set of children of $v_i$, and $incoming(v_j,v_i)$ be
the number of incoming non-tree edges ($v_l,v_i$) such that $v_l$
is a descendant of $v_j\in C_i$. Then $count_i$ is calculated
recursively as follows:

$count_i:= \underset{v_j\in C_i}{\sum}count_j -in_i+out_i;$

The algorithm is based on the Theorem~\ref{tarjan1} and
Theorem~\ref{tarjan2} due to Tarjan~\cite{tarjandfs}.
\begin{thm}\label{tarjan1}
\begin{itemize}\item [(i)] If a non-root node $v_i$ has a child $v_j$ in $T$, then
$v_i$ is an articulation point of $G$ if and only if
$count_j=incoming(v_j,v_i)$. \item [(ii)]  The root $v_1$ is an
articulation point of $G$  if and only if $v_1$ has two or more
children.\end{itemize}
\end{thm}

\begin{thm}\label{tarjan2}
Let ($v_i, v_j$) be a tree edge in $T$ such that $v_i$ is the
parent of $v_j$. Then ($v_i, v_j$) is a bridge in $G$ if and only
if $count_j=0$.
\end{thm}

Corollary~\ref{cor1} follows from Theorem~\ref{tarjan1} and
Theorem~\ref{tarjan2}.

\begin{cor}\label{cor1}
Each of the end nodes of a bridge is an articulation point unless
it is a node of degree one.
\end{cor}

\begin{remark}
For each leaf node $v_i$, $in_i=0$ and $count_i=out_i$.
\end{remark}

In order to extend this depth-first search algorithm to find the
bridges and articulation points, and bridge-connected components
every processor $v_i$, in addition to the field $path_i$,
maintains two fields: $count_i$ and $bcc_i$. The field $bcc_i$ is
a unique identifier of the bridge-connected component containing
$v_i$. For every bridge-connected component, a
\emph{\textbf{representative node}} is defined. A
\emph{\textbf{representative node}} $v_j$ of a bridge-connected
component is the ancestor of all other nodes of the component
containing $v_j$. When the algorithm stabilizes, every
bridge-connected component is uniquely identified by the
$path$-value of its \emph{representative node}, and $bcc$-fields
of all nodes of this component contain this $path$-value.

\begin{lem}\label{lemon1}
A node $v_i$ is a representative node if and only if $count_i=0$.
\end{lem}
\begin{IEEEproof}Let $v_i$ be a \emph{representative node} and
$count_i>0$. Let $(v_m, v_l)$ be a non-tree edge such that $v_l$
is an ancestor and $v_m$ is a descendant of $v_i$. Node $v_l$ can
be reached from $v_i$ using the tree path $v_i-v_m$ followed by
the non-tree edge ($v_m, v_l$) while $v_l$ can also be reached
from $v_i$ using another path $v_i-v_l$ and these two paths are
disjoint. That is, the ancestor $v_l$ is bridge-connected to $v_i$
which contradicts that $v_i$ is a \emph{representative node}.
Again, by Theorem~\ref{tarjan2}, if $ count_i=0$ then no ancestor
of $v_i$ can be reached from $v_i$ when the parent link of $v_i$
is removed. Hence $v_i$ is a \emph{representative node} of the
bridge-connected component containing $v_i$,
\end{IEEEproof}

During the execution of the algorithm, every non-root node $v_i,
i\ge 2$, repeatedly reads in $count_j$ of every $v_j\in C_i$, and
based on $in_i$  and $out_i$, it counts the value $count_i$.
Furthermore, every \emph{representative node} $v_i$ repeatedly
writes its own path value $path_i$ into $bcc_i$ field and every
\emph{non-representative node} $v_l$ repreatedly reads in $bcc_m$
of its parent $v_m$ and writes this value into $bcc_l$. The root
$v_1$ always writes 0 into $count_1$ field and $\bot$ (i.e.
$path_1$ value)  into $bcc_1$ field.

The algorithm is presented as the \textbf{2-EDGE \& 2-VERTEX
CONNECTIVITY} algorithm. The functions \textbf{\emph{read}} and
\textbf{\emph{write}} are the functions for reading from and
writing to a register, respectively. The fields in the register of
$v_i$, $1\le i\le n$, are: $path_i$, $count_i$, $bcc_i$; the local
variables are $path$, $read\_path_j$, $read\_count_j$, and
$read\_bcc_j$ ($1\le j\le \delta_j$).

\begin{thm}\label{fair}
For every fair execution of the \textbf{2-EDGE \& 2-VERTEX
CONNECTIVITY} algorithm, there is a suffix in which for every node
$v_i$, $1\le i\le n$, $bcc_i=path_t$ in every configuration, where
$v_t$, $1\le t\le n$, is the \textbf{representative node} of the
bridge-connected component containing $v_i$.
\end{thm}
\begin{IEEEproof}
In the \textbf{2-EDGE \& 2-VERTEX CONNECTIVITY} algorithm,  new
instructions for determining the bridges, and articulation points
are embedded in the self-stabilizing depth-first search algorithm
of Collin and Dolev~\cite{collin}. These new instructions do not
affect the original function of the depth-first search algorithm.
Therefore, by Theorem 3.2 in~\cite{collin}, for every fair
execution of the \textbf{3-EDGE-CONNECTIVITY} algorithm, there is
a suffix $S$ of the execution in which $path_i$, $1\le i\le n$,
contains the correct value in every configuration. Suppose the
execution has reached a configuration $c$ in $S$. By Observation
3.1 in~\cite{collin}, the correct values in $path_i$, $1\le i\le
n$, specify a depth-first search tree $T$.

 Let $v_i$ be any leaf node in $T$. Since $path_j$, $1\le i\le n$, is correctly determined,
 after $v_i$ reads in the $path$ field from each outgoing non-tree link,
 $count_i$ value is correctly determined.
 Let $S'$ be a suffix of the execution in which all the
nodes on level $h$ or higher (i.e. farther from the root) have
correctly computed their $count$ values. Consider any non-leaf
node $v_i$, on level $h-1$.  By the induction hypothesis, for each
$v_j\in C_i$, the values of $count_j$ are correctly calculated.
Therefore, $v_i$ correctly calculates $count_i$. Hence, there is a
suffix of the suffix $S'$ in which for every configuration,
$count_i$ are correctly computed for every node $v_i$, $1\le i\le
n$.

Suppose the execution has reached a configuration in the
aforementioned suffix of suffix $S'$.
 The root node $v_1$ is a \emph{representative node} and always correctly writes
 the value of $path_1$ (i.e. $\bot$) into the field $bcc_1$.
Let $S''$ be a suffix in the suffix $S'$ of the execution in
which, for every node $v_m$ on level $h$ or lower (i.e. closer to
the root),  $bcc_m=path_l$, where $v_l$ is the
\emph{representative}  node of the bridge-connected component
containing $v_m$. Let $v_i$ be any non-root node on level $h+1$.
 The value $next_i$ can be read from $S_i$ which is correctly calculated.
 If $v_i$ is a \emph{representative} node, then $v_i$
 correctly writes $path_i$ into $bcc_i$.  If $v_i$ is not a \emph{representative} node, then $v_i$
reads in the $bcc$-field of its parent which is correct by the
induction hypothesis and writes it into $bcc_i$. Hence, there is a
suffix of the fair execution in which for every node $v_i$, $1\le
i\le n$, $bcc_i=path_t$ in every configuration, where $v_t$, $1\le
t\le n$, is the \emph{representative node} of the bridge-connected
component containing $v_i$.
\end{IEEEproof}

\begin{figure*}
\centering
\includegraphics[width=0.75\textwidth]{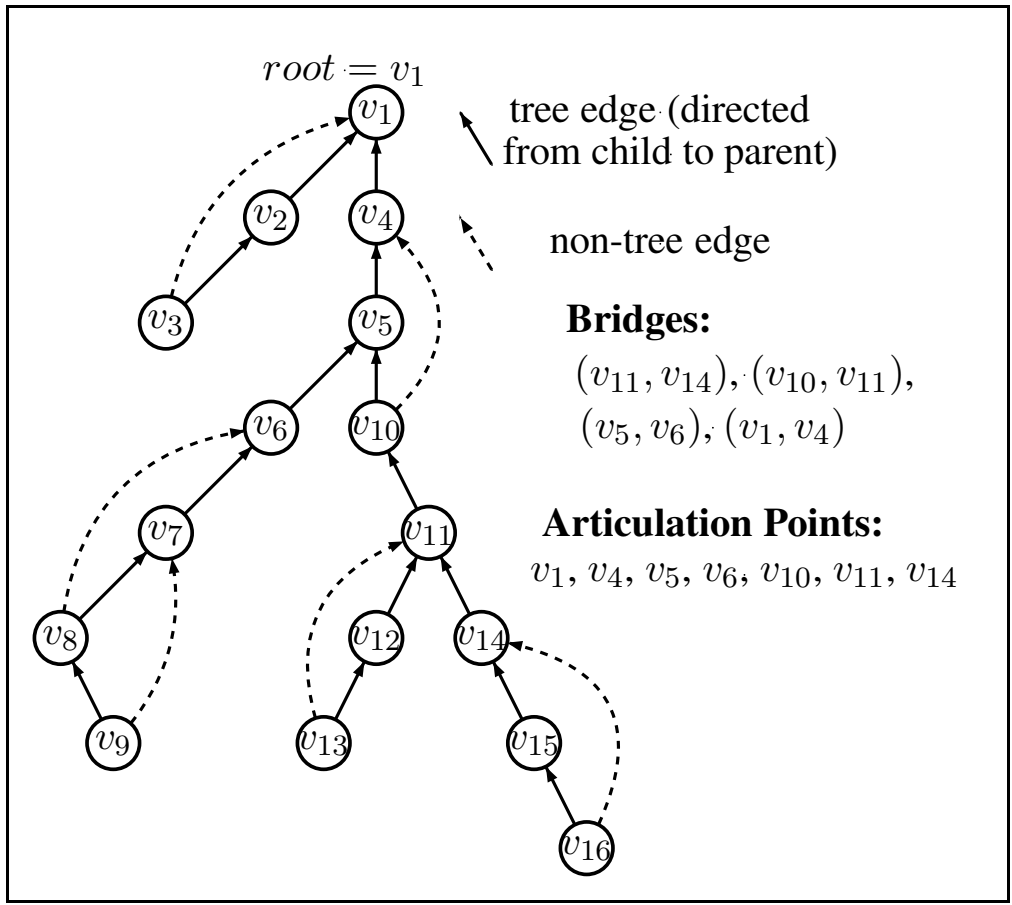} 
\caption{Depth-First Search Spanning Tree $T$}\label{mainfig}
\end{figure*}

\begin{lem}\label{determine}
When the \textbf{2-EDGE \& 2-VERTEX CONNECTIVITY} algorithm
stabilizes, all the bridges, articulation points, and
bridge-connected components are determined.
\end{lem}
\begin{IEEEproof}
When the algorithm stabilizes, by Theorem~\ref{fair}, every node
$v_i, 1\le i\le n,$ knows its children, parent, all incident
tree-edges and non-tree edges, $count_i$ values, and $bcc_i$
values. By Theorem~\ref{tarjan2}, any tree edge ($v_i, v_j$) with
$count_i=count_j=0$ is a bridge, and, by Corollary~\ref{cor1},
each of these two nodes is an articulation point unless its degree
is one. By Theorem~\ref{tarjan1}, any other non-root node $v_i$
having a child $C_j$ such that $count_j=incoming(v_j,v_i)$ is an
articulation point. If the root $v_1$ has more than one child then
$v_1$ is an articulation point. Every bridge-connected component
has a unique \emph{representative node} and, by
Theorem~\ref{fair}, the \emph{path} value of this node is written
into $bcc$ field of every node of this component. Hence $bcc_i$
value of every node $v_i$, $1\le i\le n$,  uniquely identifies the
bridge-connected component containing $v_i$.
\end{IEEEproof}

\begin{algorithm}[H]
\begin{small}
\textbf{2-EDGE \& 2-VERTEX CONNECTIVITY} \textbf{Algorithm}

Let ${v_i}_j$, $1\le j \le \delta_i$, be the neighboring
processors of processor $v_i$, $1\le i\le n$, such that
$\alpha_i(i_j)=j$, $1\le j\le \delta_i$, $1\le i\le n$.

\textbf{root} $v_1$:

\For{forever}{ \textbf{write} $path_1:=\bot$; \textbf{write}
$count_1:= 0$; \textbf{write} $bcc_1:=\bot $;}

\textbf{non-root} $v_i$, $i\ge 2$:

\For{forever}{ \tcc*[h]{Calculate $path_i$ }

\lFor{$j:=1$ to $\delta_i$}{$read\_path_j:=$
\textbf{read}(${path_i}_j$);\tcc*[f]{read path-value of neighbor
${v_i}_j$ }}

 \textbf{write} $path_i:=$\textbf{\emph{min}}$_{\prec}\{|read\_path_j\oplus{\alpha_i}_j(i)|_N$ such that $1\le
j\le \delta_i\}$; \tcc*[f]{compute $path_i$ }

\tcc*[h]{Calculate $count_i$ }

$path:=$ \textbf{read}($path_i$);

$in:=out:= count:=0$; \tcc*[f]{initialize $in, out,$ and $count$ }

\For{$j:=1$ to $\delta_i$}{
        \If(\tcc*[f]{($v_i,{v_i}_j$) is a child link}){($read\_path_j=path\oplus {\alpha_i}_j(i)$)}{
                $read\_count_j:=$\textbf{read}(${count_i}_j$);  \tcc*[f]{read $count$ value
                of child ${v_i}_j$}\\
               $count:=count+read\_count_j$;\tcc*[f]{update $count$}
               }

        \uElseIf(\tcc*[f]{incoming non-tree edge}){   ($\exists s$)(($read\_path_j=path\oplus s$) $\wedge$ ($s\not= {\alpha_i}_j(i)$))  }{ $count:=count-1$;\tcc*[f]{update $count$}}
        \uElseIf(\tcc*[f]{outgoing non-tree edge}){   ($\exists s$)(($path=read\_path_j\oplus s$) $\wedge$ ($s\not= {\alpha_i}_j(i)$))  }{ $count:=count+1$;\tcc*[f]{update $count$}}

    }

\textbf{write} $count_i:=count$; \tcc*[f]{write $count_i$ }

\tcc*[h]{Calculate bridge-connected component identifier $bcc_i$ }

$count:=$ \textbf{read}($count_i$);
$path:=$\textbf{read}($path_i$);

\uIf{$(count=0)$}{\textbf{write} $bcc_i:=path$;\tcc*[h]{$v_i$ is a
representative node. Write $path_i$ into $bcc_i$ }}
                     {\uElse{
\For{$j:=1$ to $\delta_i$}{
        \If(\tcc*[f]{($({v_i}_j, v_i$) is the parent link}){($path=read\_path_j\oplus {\alpha_i}_j(i)$)}{$read\_bcc_j:=$\textbf{read}(${bcc_i}_j$); \textbf{write} $bcc_i:=read\_bcc_j$;\tcc*[f]{(write $bcc$ of ${v_i}_j$ in $bcc_i$}}
}
                      }
                     }

} 
\end{small}
\end{algorithm}

\begin{lem}\label{time}
The \textbf{2-EDGE \& 2-VERTEX CONNECTIVITY} algorithm stabilizes
in  O($dn\Delta $) rounds, where $\Delta$ is an upper bound on the
degree of a node, $d$ is the diameter of the graph.
\end{lem}
\begin{IEEEproof}
It is easily verified that the new instructions added to the
depth-first search algorithm of Collin and Dolev~\cite{collin}
only increase the time complexity for constructing a depth-first
search tree by a constant factor.  The \textbf{for} loop for
computing the $count$-values takes $O(H)$ rounds, where $H$ is the
height of $T$ and  the \textbf{for} loop for computing the $bcc$
values takes O(1)
 rounds. Therefore, the time required by
the \textbf{2-EDGE \& 2-VERTEX CONNECTIVITY} algorithm remains
same as that of the underlying depth-first search algorithm (i.e.
O($dn\Delta$) rounds).
\end{IEEEproof}

\begin{lem}\label{space}
The space complexity of the  \textbf{2-EDGE \& 2-VERTEX
CONNECTIVITY} algorithm is O($n\log\Delta$) bits per processor.
\end{lem}
\begin{IEEEproof}
In the depth-first search algorithm of Collin and
Dolev~\cite{collin}, the space required by every processor is
O($n\log\Delta$) bits. This is the space required to store the
\emph{path} value of the processor. In the \textbf{2-EDGE \&
2-VERTEX CONNECTIVITY} algorithm, $bcc$ field requires
O($n\log\Delta$) bits, and $count$ filed requires O($\log
(n\Delta)$) $\approx $ O($\log n$) bits. The space complexity per
processor is thus O($n\log\Delta$) bits.
\end{IEEEproof}

Figure~\ref{mainfig} is a depth-first spanning tree of the
corresponding undirected graph. An execution of our algorithm over
this tree is shown below.

The $count$ values at non-root nodes $v_4$, $v_6$, $v_{11}$, and
$v_{14}$ are 0, and hence, by Lemma~\ref{determine}, the bridges
are $(v_{11} , v_{14})$, $(v_{10}, v_{11})$, $(v_5 , v_6)$, $(v_1,
v_4)$. The articulation points are $v_1$, $v_4$, $v_5$, $v_6$,
$v_{10}$, $v_{11}$, $v_{14}$, and bridge-connected components are:
$\{v_1, v_2, v_3\}$, $\{v_4, v_5, v_{10}\}$, $\{v_6, v_7, v_8,
v_9\}$, $\{v_{11}, v_{12}, v_{13}\}$, and $\{v_{14}, v_{15},
v_{16}\}$.

\section{Conclusion}\label{conclusion}
We have presented an algorithm for the 2-edge-connectivity and
2-vertex-connectivity problem based on a self-stabilizing
depth-first search algorithm. The algorithm constructs a
depth-first search tree in $O(dn\Delta)$ rounds and then
determines the bridges, articulation points, and bridges-connected
components based on the depth-first search tree. In the worst
case, when $d= \Delta = n$, our algorithm requires $O(n^3)$
rounds. Clearly, the time complexity of our algorithm is dominated
by the time spent in constructing the depth-first search tree.
Should there be an improvement made on the time bound required to
construct the depth-first search tree, the time complexity of our
algorithm will improve as well.

\bibliographystyle{plain}
\bibliography{biblio}

\begin{thebibliography}{10}

\bibitem{chaudhurybcc}
P.~Chaudhuri.
\newblock An \emph{O}$(n^2)$ self-stabilizing algorithm for computing
  bridge-connected components.
\newblock {\em Computing}, 62(1):55--67, February 1999.

\bibitem{chaudhuryfc}
P.~Chaudhuri.
\newblock A self-stabilizing algorithm for detecting fundamental cycles in a
  graph.
\newblock {\em Journal of Computer and System Science}, 59(1):84--93, August
  1999.

\bibitem{collin}
Zeev Collin and Shlomi Dolev.
\newblock Self-stabilizing depth-first search.
\newblock {\em Information Processing Letters}, 49(6):297--301, March 1994.

\bibitem{devismes}
St$\acute{e}$phane Devismes.
\newblock A silent self-stabilizing algorithm for finding cut-nodes and
  bridges.
\newblock {\em Parallel Processing Letters}, 15(1\&2):183--198, March \& June
  2005.

\bibitem{dijkstra}
Edsger~W. Dijkstra.
\newblock Self-stabilizing systems in spite of distributed control.
\newblock {\em Communications of the ACM}, 17(1):643--644, November 1974.

\bibitem{belated}
Edsger~W. Dijkstra.
\newblock A belated proof of self-stabilization.
\newblock {\em Distributed Computing}, 1(1):5--6, January 1986.

\bibitem{dolev00}
S~Dolev.
\newblock {\em Self-stabilization}.
\newblock MIT Press, Cambridge, Massachusetts, 2000.

\bibitem{karaataarticulation}
M.~H. Karaata.
\newblock A self-stabilizing algorithm for finding articulation points.
\newblock {\em International Journal of Foundations of Computer Sciences},
  10(1):33--46, 1999.

\bibitem{karaatabc}
M.~H. karaata.
\newblock A stabilizing algorithm for finding biconnected components.
\newblock {\em Journal of Parallel and Distributed Computing}, 62(5):982--999,
  May 2002.

\bibitem{karaatabridge}
Mehmet~Hakan Karaata and Pranay Chaudhuri.
\newblock A self-stabilizing algorithm for bridge finding.
\newblock {\em Distributed Computing}, 12(1):47--53, March 1999.

\bibitem{ispa}
Abusayeed~M. Saifullah and Yung~H. Tsin.
\newblock A self-stabilizing algorithm for 3-edge-connectivity.
\newblock In {\em Proceedings of the 5th International Conference on Parallel
  and Distributed Processing and Applications}, ISPA'07, pages 6--19, 2007.

\bibitem{ijfcs}
Abusayeed~M. Saifullah and Yung~H. Tsin.
\newblock Self-stabilizing computation of 3-edge-connected components.
\newblock {\em International Journal of Foundations of Computer Science},
  22(05):1161--1185, 2011.

\bibitem{cats}
Abusayeed~M. Saifullah and Alper \"{U}ng\"{o}r.
\newblock A simple algorithm for triconnectivity of a multigraph.
\newblock In {\em Proceedings of the Fifteenth Australasian Symposium on
  Computing: The Australasian Theory - Volume 94}, CATS '09, pages 53--62,
  2009.

\bibitem{tarjandfs}
R.~E. Tarjan.
\newblock Depth-first search and linear graph algorithms.
\newblock {\em SIAM J. Computing}, 1:146--160, 1972.

\bibitem{tsinss}
Y~H Tsin.
\newblock An improved self-stabilizing algorithm for biconnectivity and
  bridge-connectivity.
\newblock {\em Information Processing Letters}, 102:27--34, 2007.

\end{thebibliography}

\end{document}